# A Hybrid Parallelization of AIM for Multi-Core Clusters: Implementation Details and Benchmark Results on Ranger

Fangzhou Wei and Ali E. Yılmaz

**Abstract**

This paper presents implementation details and empirical results for a hybrid message passing and shared memory paralleliziation of the adaptive integral method (AIM). AIM is implemented on a (near) petaflop supercomputing cluster of quad-core processors and its accuracy, complexity, and scalability are investigated by solving benchmark scattering problems. The timing and speedup results on up to 1024 processors show that the hybrid MPI/OpenMP parallelization of AIM exhibits better strong scalability (fixed problem size speedup) than pure MPI parallelization of it when multiple cores are used on each processor.

# 1 INTRODUCTION

This paper empirically investigates the parallel scalability of FFT accelerated iterative method of moments (MOM) solvers on clusters of multi-core processors. It complements [1], which presents in detail the parallelization of both the classical and the adaptive integral method (AIM) [2-4] accelerated iterative MOM solution of integral equations pertinent to the analysis of time-harmonic electromagnetic scattering from perfect electrically conducting (PEC) objects. The reader is referred to [1] for a review of the typical pure message-passing (MPI) based parallelization approaches, hybrid message-passing/shared-memory (MPI/OpenMP) based techniques, and general complexity and scalability expressions for MOM and AIM.

This paper demonstrates the performance and multi-core scalability of the MPI/OpenMP-AIM scheme in practice. It implements the methods described in [1] and applies them to several benchmark electromagnetic scattering problems that represent best-case, worst-case, and complex-application scenarios. It presents comprehensive accuracy, performance, and scalability results for MPI/OpenMP-AIM and contrasts them to those for MPI-AIM on the state-of-the-art Ranger cluster [5] for problems with up to $N \sim 10^7$ degrees of freedom. The benchmark scattering problems are introduced in Section 2, the performance of parallel MOM and AIM are quantified for them in Sections 3 and 4, and conclusions are presented in Section 5.

# 2 BENCHMARK SCATTERING PROBLEMS

This section details the scattering simulations that are used throughout the paper to investigate computational costs. First, three different benchmark scatterers are described. Second, target error levels are identified for the simulations, the MOM and AIM parameters for achieving these levels are specified, and the errors resulting from the actual simulations are listed. Third, key properties of the Ranger cluster are identified and the expected performances of the methods are presented. Fourth and finally, the computational complexity of the implementations are verified.

## 2.1 Plate, Sphere, and Model Airplane

Three PEC scatterers are used for benchmarking: A square plate of $L_\text{P} = 1 \text{ m}$ side length, a sphere of radius $L_\text{S} = 1 \text{ m}$, and a model airplane of $L_\text{A} = 0.82 \text{ m}$ nose-to-tail length that fits into a rectangular prism of size $\sim 0.82\text{m} \times 0.81\text{m} \times 0.24\text{m}$ [4], [6]. The plate resides on the x-y plane centered at origin; the sphere is centered at origin; and the model airplane's wings are parallel to the *x-y* plane with the nose pointing in the $\hat{y}$ direction. The plate and sphere are illuminated by an $\hat{x}$ polarized plane wave propagating toward $-\hat{z}$ direction; the model airplane is illuminated by a $\hat{z}$ or $-\hat{x}$ (vertical or horizontal) polarized plane wave propagating toward $-\hat{y}$ direction. In the following, increasingly larger simulations are performed by repeatedly doubling the frequency of interest and keeping the surface mesh density constant with respect to wavelength (all scatterer surfaces are meshed using triangular elements with average edge lengths of approximately $\lambda/9$ for the plate and sphere and $\lambda/11$ for the model airplane). The plate is simulated for 9 frequencies from 0.3 GHz to 76.8 GHz; its side length ranges from $\lambda$ to $256\lambda$, where $\lambda$ denotes the wavelength of interest. The sphere is simulated for 7 frequencies from 0.3 GHz to 19.2 GHz; its radius ranges from $\lambda$ to $64\lambda$. The model airplane is simulated for 5 frequencies from 2.5 GHz to 40 GHz (each model mesh is obtained by uniformly refining the original triangular mesh); its nose-to-tail length ranges from $\sim 6.8\lambda$ to $\sim 108\lambda$, i.e., the model fits into rectangular prism of approximately $6.8\lambda \times 6.7\lambda \times 2\lambda$ and $108\lambda \times 106\lambda \times 32\lambda$ size for the lowest and highest frequencies, respectively.

The plate and sphere represent the best- and worst-case extremes for AIM as the number of regular grid points $N^C$ is proportional to the number of surface unknowns $N$ for plates and to $N^{1.5}$ for spheres [1-4]. The model airplane represents a complex-application scenario that demonstrates the method's versatility.

**2.2 Error Targets, Simulation Parameters, and Observed Errors**

The accuracy of the simulations are quantified by computing the relative root-mean-square error in the VV-polarized bistatic radar cross section (RCS) $\sigma_{\theta\theta}$, which is denoted by $err_{\theta\theta}$:

$$err_{\theta\theta} = \left( \frac{\int_0^{2\pi}\int_0^{\pi} \left|\sigma_{\theta\theta}^{\mathrm{AIM}} - \sigma_{\theta\theta}^{\mathrm{ref}}\right|^2 \sin\theta d\theta d\phi}{\int_0^{2\pi}\int_0^{\pi} \left|\sigma_{\theta\theta}^{\mathrm{ref}}\right|^2 \sin\theta d\theta d\phi} \right)^{1/2} \quad (1)$$

The reference RCS results for the above error measure are chosen as follows. For plate and model airplane, because exact results do not exist, approximate but more accurate simulation results are used as reference: The classical MOM solution of the same problem is used whenever possible; when this is not feasible because of high computational cost, a more accurate (less efficient) AIM solution, which is obtained by increasing the moment matching order, is used as reference (up to fifth order moments are matched to obtain the reference results). For sphere simulations, the analytical Mie series solutions are used as the primary reference; moreover, to gauge their usefulness as references for the other scatterers, the MOM and the more accurate AIM solutions are also used as references. The AIM parameters are chosen empirically to minimize the computational costs subject to the following error constraints: When analytical results are used as reference, $err_{\theta\theta} < 1\%$; when numerical results are used as reference, $err_{\theta\theta} < 0.5\%$ (out of an abundance of caution). The number of unknowns $N$ and the number of regular grid points $N^C$ are listed in Tables 1-3 for all the benchmark simulations (up to third order moments are matched and $\gamma$, the parameter that sets the near-zone correction size [1], is 3 in all cases). The tables show that the more accurate AIM simulations serve as a reference equivalent to classical MOM simulations (when they are feasible) and that the observed errors satisfy the dictated constraints. It should be noted that the electric-field integral equation formulation [1] is used for plate simulations because it is an open scatterer and the combined-field integral equation formulation (with linear combination parameter $\alpha = 0.6$) [1] is used for sphere and model airplane simulations. A diagonal pre-conditioner is used in all simulations and the iterative solver is terminated when the relative root-mean-square error of the residual is less than $10^{-4}$.

To demonstrate the accuracy of the AIM simulations, Figs. 1(a)-(b) compare the VV-polarized bistatic RCS results for the highest-frequency plate and sphere simulations to the approximate physical optics [7] and exact Mie series results, respectively. Fig. 2 compares the VV- and HH-polarized bistatic RCS results for the model aircraft at 1.5 GHz with the reference MOM simulation results in [6], [8]. In all cases, the observed agreements with reference results are as expected. As an additional verification, Fig. 3 compares the VV-polarized monostatic RCS for the model aircraft at 10 GHz with measured data [9], [10]. Even though there is a high correlation between the simulated and measured RCS data, there are significant disagreements. The discrepancies are primarily due to the inaccurate geometry model: The model airplane mesh for the 10 GHz simulations were generated by refining the mesh that was generated for the 1.5 GHz simulations. While this method yields a mesh that resolves the smaller wavelength, it does not improve the resolution of the surface curvature and other geometrical features beyond that of the original mesh suitable for lower frequencies.

**2.3 Expected Complexity and Scalability**

The simulations in this paper are conducted on Ranger [5], which is a near-petaflop cluster of 3936 computing nodes consisting of four quad-core processors and 8 GB of memory per processor (2 GB per core); thus, the number of active processors $M$ and the number of active cores per processor $T$ are bounded as $M \leq 15744$ and $T \leq 4$. The time for computing one floating point operation, the latency, and the inverse of the network bandwidth on Ranger are approximately $t_{\text{fl}} \sim 0.45 \text{ ns}$, $t_{\text{lat}} \sim 4.5 \text{ }\mu s$, and $t_{\text{bw}} \sim 1 \text{ ns / byte}$, respectively.

The MOM matrix fill time, memory requirement, and matrix solve time per iteration are expected to scale as $O(N^2)$ irrespective of the scatterer. On Ranger, the scalability of the MOM matrix solve time is expected to be latency limited for both MPI-MOM because $t^{\text{bw}} < \sqrt{t^{\text{fl}} t^{\text{lat}} T}$ and MPI/OpenMP-MOM because $t^{\text{bw}} < \sqrt{t^{\text{fl}} t^{\text{lat}} / T}$ for all $T \leq 4$. As a result, MPI/OpenMP-MOM should reduce $t_{\min}^{\text{solve}}$, the minimum time that can be achieved per iteration, by a factor of $T$ as compared to MPI-MOM by using $P_{\max}^{\text{solve}} T$ instead of $P_{\max}^{\text{solve}}$ cores, irrespective of the scatterer [1].

The complexity and scalability of AIM depend on the scatterer: (i) For the plate, AIM matrix fill time, memory requirement, and matrix solve time per iteration are expected to scale as $O(N)$, $O(N)$, and $O(N \log N)$, respectively. On Ranger, the scalability of the matrix solve time is expected to be latency limited for both MPI-AIM and MPI/OpenMP-AIM because $P_1^{\text{lat}} T < P_{\max}^{\text{FFT}}$ according to the definitions in [1]. (ii) For the sphere, AIM matrix fill time, memory requirement, and matrix solve time per iteration are expected to scale as $O(N)$, $O(N^{1.5})$, and $O(N^{1.5} \log N)$, respectively. On Ranger, the scalability of the AIM matrix solve time could be latency limited for the smaller sphere simulations but is expected to be grid limited for the larger ones for both parallelization schemes [1]. (iii) For the model airplane, AIM matrix fill time, memory requirement, and matrix solve time per iteration are expected to scale as $O(N)$, $O(N^{1.5})$, and $O(N^{1.5} \log N)$, respectively; these expressions are similar to those for the sphere, but the constants in front of the complexity estimates should be approximately one order of magnitude smaller. Thus, the scalability of the AIM matrix solve time for the model airplane will also be grid limited eventually; however, it is expected to remain latency limited for simulations with up to an order of magnitude larger number of unknowns compared to the sphere simulations.

## 2.4 Computational Complexity Validation

Fig. 4 shows the computational requirements of the classical and AIM accelerated MOM solvers observed on the Ranger cluster. These data are obtained from the hybrid MPI/OpenMP parallel implementation of the methods detailed in Sections 3 and 4; specifically, the timing data are obtained by multiplying the observed wall-clock times with the number of active cores and the memory data are obtained by summing the memory required by all cores. To minimize the effect of parallelization inefficiencies, the data in Fig. 4 are obtained by using only the minimum number of cores dictated by the memory requirements and only one core per processor. The results observed in Fig. 4 agree well with the theoretically expected computational complexity values. Notice that AIM accelerated MOM solvers outperform classical solvers in all performance metrics when $N$ is as small as $10^3$. Several irregularities are evident in the observed data: (i) In Fig. 4(a), the matrix fill time for the model airplane scales as $O(N)$ yet with a larger constant in front of the complexity estimate compared to the sphere. This is due to the slightly smaller average edge length of the model airplane mesh as well as the thin structures in the wings and the tail, which lead to a larger near zone correction region compared to the sphere and a higher matrix fill time. (ii) In Fig. 4(b), the memory requirements slightly deviate from the computational complexity line for the largest two plate simulations. This is because of parallelization inefficiencies (specifically, non-parallelized data replication) and is explained in detail in Section 4.2. (iii) In Fig. 4(b), the total memory requirement for the sphere and model airplane scale as $O(N)$ when $N$ is relatively small ($N < 10^5$) but as $O(N^{1.5})$ when $N$ is larger. This is because the $\mathbf{Z}^{\text{near}}$ matrix initially dominates the memory cost while the $\mathbf{Z}^{\text{FFT}}$ matrix eventually becomes the major memory cost as the size of the sphere increases [1]. (iv) In Fig. 4(c), a jump in the solution time is observed for the $L_s = 32\lambda$ sphere case. A closer investigation of the data shows that this jump is caused by a disproportionate increase in the time needed for calculating FFTs during the AIM propagation step: In theory, the time needed for the $N^C = 512^3$ auxiliary grid (for which $1024$ 2-D FFTs of size $1024 \times 1024$ are calculated) should be ~9 times larger than that needed for calculating FFTs for an $N^C = 256^3$ grid (for which $512$ 2-D FFTs of size $512 \times 512$ are calculated); but the observed time was ~17 times larger. For smaller and larger problem sizes, the FFT time (as well as the total solution time in Fig. 4(c)) was observed to scale as expected; thus, this jump likely indicates the point where the large FFT arrays start to not fit in the faster but limited cache memory.

# 3 PARALLEL MOM RESULTS

This section presents numerical results that show the parallel scalability of the MPI-MOM and MPI/OpenMP-MOM for the plate; similar results were obtained for the other benchmark scatterers [11] but are not shown here for expediency. In the following, the wall-clock time required for the matrix fill step, the memory cost, and the average wall-clock time required for one iteration during the matrix solve step are measured while varying $M$, the number of active multi-core processors, and $T$, the number of active cores in each processor.

Fig. 5 shows the computational requirements for several benchmark plate simulations along the (logarithmic) $y$ axis versus the total number of active cores $P = MT$ along the (logarithmic) $x$ axis, respectively. Multiple values of $M$ and $T$ can result in the same total number of active cores; to avoid any confusion, the data in the figures are plotted by fixing $T$ and varying $M$, Figs. 5(a), (c), and (e) (Figs. 5(b), (d), and (f)) show the wall-clock time during the matrix fill step, the maximum memory required per core[1] during the entire analysis, and the average wall-clock time for one iteration during matrix solve step required by the MPI-MOM (hybrid MPI/OpenMP-MOM) implementation for the different plate simulations, respectively. Figs. 5(a)-(b) show that the matrix fill step of both MOM implementations exhibit near-ideal scalability. Moreover, the lines for different choices of $T$ coincide, i.e., the time required for the matrix fill step does not depend on the particular choice of $M$ and $T$ but is a function of the total number of cores $P = MT$. Figs. 5(c)-(d) show that the maximum memory requirement of the hybrid MPI/OpenMP-MOM implementation exhibits better scalability. The memory requirement of both implementations falls down to a constant level as the total number of active cores increases; this level is dictated by the operating system, compiler, and parallel library overheads as well as the storage requirements of the non-parallelized geometry and basis/testing function data that are replicated among all MPI processes. The MPI/OpenMP-MOM memory requirement is practically independent of $T$ as expected. Note that the memory requirement of the MPI-MOM begins to increase slowly with the number of cores after scaling down to 100 MB for the smallest plate case; this is due to the overhead of auxiliary data structures and operations specifically used for parallelization that scale with the number of cores (but not with the number of unknowns). This overhead is due to both the MPI library and the MPI-MOM implementation. The memory requirement of the hybrid MPI/OpenMP-MOM will also suffer from this problem as the number of MPI processes increases. Figs. 5(e)-(f) show that the matrix solve step of the hybrid MPI/OpenMP-MOM exhibits better scalability and is less sensitive to the choice of $T$. It is observed that the fewer number of cores are active in each processor (the smaller $T$ is), the more scalable MPI-MOM is with $P$; in comparison, the hybrid MPI/OpenMP-MOM scales almost independently of the number of active cores on each processor and it effectively uses $T$ times more cores to reduce the minimum matrix solve time by the factor of $T$. These results are in agreement with the analysis in [1] (compare Fig. 5 to Fig. 7 of [1]); moreover, it is evident that both implementations are latency limited on Ranger as no plateau regions are observed around the turning points. Lastly, it should be observed that all computational requirements of both MOM implementations exhibit weak scalability, i.e., the larger the problem size the more cores can be used effectively; in line with the prediction that $P_{\max}^{\text{solve}}$ is proportional to $N$ [1].

---

[1] "The maximum memory required per core" is found by calculating the maximum of the following data: The peak memory requirement of each MPI process measured during run time for the MPI-MOM implementation; and the peak memory requirement of each MPI process measured during run time divided by the number of OpenMP threads for the hybrid MPI/OpenMP-MOM implementation. This is a convenient measure of memory use on multi-core processors that captures any memory imbalance among MPI processes but does not account for the memory imbalance among threads that share the same memory space.

## 4 PARALLEL AIM RESULTS

This section presents numerical results that compare the parallel scalability of MPI-AIM and MPI/OpenMP-AIM on Ranger. First, two possible 3-D FFT implementations are contrasted; then, using the more efficient one, the benchmark scatterers are simulated.

### 4.1 Implementation Choices: Blocking vs. Non-blocking 3-D FFTs

Both message passing and hybrid message passing/shared memory parallelization of AIM requires a global matrix transpose during the 3-D forward and inverse FFTs at the propagation step (see section 4 in [1]); this transpose can be implemented by using either a (blocking) collective communication or a (non-blocking) point-to-point communication approach [12]. These are detailed and contrasted next; to simplify the presentation, only the forward FFTs are described, only the MPI-AIM parallelization is considered, and it is assumed that the number of active cores $P = MT$ is a divisor of both $N^{cx}$ and $N^{cy}$, the number of auxiliary grid points along $x$ and $y$ dimensions [1]. The following analysis remains valid for inverse FFTs, the MPI/OpenMP-AIM parallelization, and when $P$ is not a divisor.

Consider the matrix transpose during the 3-D FFTs at the propagation step (Fig. 6(a)): Each MPI processes is assigned $2N^{cx}/P$ columns (a column slab) of the auxiliary grid before the matrix transpose and $2N^{cy}/P$ rows (a row slab) of the auxiliary grid after the matrix transpose; the number of grid points in the slab assigned to each process is $(4N^{cy}N^{cz})2N^{cx}/P$ before and $(4N^{cx}N^{cz})2N^{cy}/P$ after the transpose. Thus, each matrix transpose requires an all-to-all communication where each process receives and sends $O(N^C/P)$ bytes of data.

In the collective communication approach, the processes communicate only after all of them finish computing the 2-D FFTs along the *y-z* dimensions in their column slabs and no process can compute the 1-D FFTs along the *x* dimension in their row slabs until all of them have finished sending (and receiving) data. This approach relies on the "MPI_Alltoall" collective communication directive and is very commonly used, e.g., the multi-dimensional FFT subroutines in the FFTW library adopt this approach [13]. In the point-to-point communication approach, first each process issues $P-1$ non-blocking "MPI_IRecv" directives to collect the row slab data. The row slab data is received in small blocks: Let $R_i C_j$ denote the block of grid points at the intersection of the $i^{\text{th}}$ row and $j^{\text{th}}$ column slab; each block contains $2N^{cz}(4N^{cx}N^{cy}/P^2)$ grid points; and the process $p$ receives the blocks $R_p C_1$ to $R_p C_{p-1}$ and $R_p C_{p+1}$ to $R_p C_P$ from the corresponding processes. Second, each process computes the 2-D FFTs in its column slab and organizes the columns into small blocks. Third, each process issues $P-1$ non-blocking "MPI_ISend" directives to distribute the column slab data, i.e., process $p$ sends the blocks $R_1 C_p$ to $R_{p-1} C_p$ and $R_{p+1} C_p$ to $R_P C_p$ to the corresponding processes. Fourth and finally, each process waits for the receive directives to be completed using "MPI_Wait" and then computes the 1-D FFTs in its row slab. Notice that computation and communication are overlapped in this approach. More importantly, the approach can be specialized to the FFTs in AIM. Specifically, because of the doubled auxiliary grid, only about half of the processes must compute the 2-D forward and 1-D inverse FFTs; this implies that the number of messages and the total data size communicated can be reduced by a factor of ~2 as only about half the processes must send data after the forward FFTs and receive data before the inverse FFTs[2].

---

[2] For the forward (inverse) FFTs: The output (input) of the AIM projection (interpolation) step is non-zero current (field) values only at the grid points immediately surrounding the scatterer; thus, only the processes that are assigned the column slabs that contain the scatterer, e.g., $P_1 - P_3$ in Fig. 6, must compute 2-D forward FFTs (1-D inverse FFTs), as the remaining processes, e.g., $P_4 - P_6$ in Fig. 6, can avoid the FFT computations because the results are zero (are not interpolated back onto the scatterer mesh).

Fig. 7 compares the performance of collective and point-to-point communication approaches for the matrix transpose at the AIM propagation step. The figure shows the average wall clock time per iteration required during matrix solve step of MPI-AIM for sample plate, sphere, and model airplane simulations. The non-blocking FFT implementation shows reduced timing and slightly better scalability for all scatterers and is adopted henceforth.

**4.2 Scalability**

Next, the parallel scalability of the two AIM parallelization techniques are contrasted; just as in Section 3, the computation requirements are measured while varying $M$, the number of active multi-core processors, and $T$, the number of active cores in each processor. Both strong and weak scalability are investigated for the two implementations.

Fig. 8 shows the strong scalability of computational requirements for several of the benchmark plates versus the total number of active cores $P = MT$. The results for other benchmark scatterers can be found in [11]. Figs. 8(a), (c), and (e) (Figs. 8(b), (d), and (f)) show the wall-clock time during the matrix fill step, the maximum memory among all cores during the entire analysis, and the average wall-clock time for one iteration during matrix solve step required by the MPI-AIM (hybrid MPI/OpenMP-AIM) implementation for different plate simulations, respectively. Figs. 8(a)-(b) show that the matrix fill step of both implementations exhibit near-ideal scalability for plates. Just like those for the MOM simulations, the AIM matrix fill time exhibits near-ideal scalability for both parallelization approaches. Figs. 8(c)-(d) show that the maximum memory requirement of the hybrid MPI/OpenMP-AIM implementation exhibits better scalability and that the hybrid parallelization reduces the minimum memory requirement per processor by a factor of $T$ compared to the distributed memory parallelization. The memory requirement of both implementations falls down to a constant level as the total number of cores increases due to the same reasons as in Section 3. Figs. 8(e)-(f) show that the matrix solve step of the hybrid MPI/OpenMP-AIM is less sensitive to the choice of $T$. Indeed, the more cores are active in each processor (the larger $T$ is) the less scalable MPI-AIM becomes. The scalability of matrix solve time is observed to be latency limited for MPI-AIM but grid limited for MPI/OpenMP-AIM. In other words, the memory and communication hierarchy of Ranger cannot be ignored for the AIM matrix solve step, and the hybrid MPI/OpenMP-AIM better exploits this hierarchy to outperform the MPI-AIM during the matrix solve step. Moreover, both parallelization approaches show limited scalability compared to their MOM counterparts in Figs. 5(c)-(d) for the same number of unknowns, as expected.

Next, the weak scalability of the two implementations are compared by focusing on the matrix solve step, which is the main parallel scalability bottleneck. Fig. 9 shows the average wall-clock time required per iteration for the benchmark scatterers; only two extreme cases are considered: $T=1$ (1 core is active per processor) and $T=4$ (all cores are active). Figs. 9(a), (c), and (e) (Figs. 9(b), (d), and (f)) show the results of all plate, sphere, and model airplane simulations for $T=1$ ($T=4$), respectively.

For the $T=1$ case, the timing data for both implementations should theoretically be identical but the data in Figs. 9(a), (c), and (e) show small differences, which are less than 5% in all cases except for the largest plate data (where the differences are less than 15%). The small differences might reflect random variations in execution time from one simulation to the next or OpenMP overheads for executing one-threaded parallel regions.

For the $T=4$ case, the advantages of the hybrid parallelization are clearly visible. Figs. 9(b), (d), and (f) show that for MPI-AIM, the matrix solve time decreases to a minimum level with $MT$, then increases and essentially becomes constant over a certain $MT$ for plate simulations, which implies that scalability is latency limited (compare Figs. 9(b), (d), and (f) to Fig. 12(d) of [1]). Similar observations can be made for small sphere and airplane simulations; however, the scalability limitation is observed to migrate from latency limited to grid limited, i.e., the constant level closes to the minimum level when the problem size increases for the sphere and model airplane. It is also observed that the matrix solve time for model airplane simulations remain latency limited for an order of magnitude larger number of unknowns than for the sphere simulations, consistent with the fact that the plate and sphere represent the best- and worst-case extremes for AIM and the model airplane simulations should fall in between the two. For the MPI/OpenMP-AIM, the matrix solve time decreases to a minimum level with $MT$ and converges to a constant value for all three benchmark scatterers, which implies that the simulations are grid limited (compare Figs. 9(b), (d), and (f) to Fig. 12(c) of [1]). In other words, the scalability is improved from latency limited for the MPI-AIM to near grid limited for the MPI/OpenMP-AIM. Moreover, the MPI/OpenMP-AIM is observed to be able to use $T$ times more cores to reduce the minimum matrix solve time for a fixed problem compared to the MPI-AIM and the minimum matrix solve time is reduced by a factor of more than 1 but less than $T$ as expected.

## 5 CONCLUSIONS

This paper demonstrated the practical performance and multi-core scalability of hybrid shared memory/message passing parallelization schemes compared to conventional message passing ones for FFT accelerated iterative MOM solvers,. The methods described in [1] were implemented and applied to three different benchmark scatterers: a plate, a sphere, and a model airplane that represent the best-case, worst-case, and complex-application scenarios for the method. The complexity and scalability of the implementations were validated for these scatters while meeting accuracy constraints. Simulations were conducted on the state-of-the-art Ranger cluster for problems with up to $N \sim 10^7$ degrees of freedom on upto $MT = 4096$ cores.

By measuring the strong as well as the weak scalability of both parallelization approaches for the benchmark scatterers, it was demonstrated that the hybrid parallelization approach is useful for alleviating memory and communication limitations as theoretically shown in [1]. As the performance improvements are a function of the number of active cores in a processor, the hybrid parallelization methods are expected to become more important as the general trend of increasing number of cores in multi- and many-core processors continues.

**Acknowledgements:** This work was supported in part by the NSF Grants ECCS-0725729 and OCI-0904907. The authors thank Prof. T. Rylander for sharing the aircraft mesh, which was originally constructed by AerotechTelub according to a design provided by the Swedish Defense Research Agency. The authors acknowledge the Texas Advanced Computing Center (TACC) at the University of Texas at Austin for providing HP resources that contributed to the research results reported within this paper.

# REFERENCES

[1] F. Wei, A.E. Yılmaz, A hybrid message passing/shared memory parallelization of the adaptive integral method for multi-core clusters, Parallel Comput., (Oct. 2010) submitted for publication.

[2] E. Bleszynski, M. Bleszynski, T. Jaroszewicz, AIM: Adaptive integral method for solving large-scale electromagnetic scattering and radiation problems, Radio Sci., 31 (1996) 1225-1251.

[3] H.T. Anastassiu, M. Smelyanskiy, S. Bindiganavale, J.L. Volakis, Scattering from relatively flat surfaces using the adaptive integral method, Radio Sci., 33 (1998) 7-16.

[4] A.E. Yılmaz, J.M. Jin, E. Michielssen, Time domain adaptive integral method for surface integral eqautions, IEEE Trans. Antennas Propagat., 52 (2004) 2692-2708.

[5] Top 500 Supercomputers: Ranger, <http://www.top500.org/system/9257/>.

[6] F. Edelvik, G. Ledfelt, Explicit hybrid time domain solver for the Maxwell equations in 3-D, J. Sci. Comput., 15 (2000).

[7] C.A. Balanis, Advanced Engineering Electromagnetics, Wiley, New York, NY, 1989.

[8] U. Andersson, Time-Domain Methods for the Maxwell Equations, Ph. D. thesis, Dep. of Numeric. Anal. & Comput. Sci., KTH, Stockholm, Sweden, 2001.

[9] N. Stefan, R. Jonas, G. Magnus, H. Magnus, K. Mikael, Z. Erik, Ö. Anders, Modeling of goals and background in the radar field, FOI-R--2359-SE, FOI, Dec. 2007.

[10] J. Löthegård, J. Rahm, N. Gustafsson, O. Lundén, C. Larsson, J. Rasmusson, K. Brage, C.-G. Svensson, Jan-Olof, Olsson, An interlaboratory comparison between the RCS ranges at FOA Defence Research Establishment and Saab Dynamics, in: Proc. AMTA, 1999, pp. 79-84.

[11] F. Wei, A hybrid MPI/OpenMP parallelization of the adaptive integral method for multi-core clusters, M.S. thesis, Dep. of Elec. & Comp. Eng., Univ. of Texas, Austin, TX, 2009.

[12] A. Dubey, D. Tessera, Redistribution strategies for portable parallel FFT: a case study, Concur. and Computat.: Prac. & Exper., 13 (2001) 209-220.

[13] M. Frigo, S. Johnson, FFTW: An adaptive software architecture for the FFT, in: Proc. Int. Conf. Acous. Speech Sig. Process., 1998, pp. 1381-1384.

**Figure Captions**

Figure 1: Bistatic RCS (VV) of the largest benchmark simulations in the *x-z* plane: (a) Plate $\left(L_\mathrm{P} = 256\lambda\right)$. (b) Sphere $\left(L_\mathrm{S} = 64\lambda\right)$.

Figure 2: Bistatic RCS (HH and VV) of the model aircraft at 1.5 GHz in the *x-y* plane.

Figure 3: Monostatic RCS (VV) of the model aircraft at 10 GHz in the *x-y* plane.

Figure 4: AIM vs. MOM for the benchmark scatterers as the frequency is increased. (a) Matrix fill time. (b) Memory requirement. (c) Average solution time per iteration.

Figure 5: Computational requirements for the $L_\mathrm{P} = 2\lambda$, $L_\mathrm{P} = 4\lambda$, and $L_\mathrm{P} = 8\lambda$ plate simulations using the two parallelization approaches. Wall clock time for matrix fill step for (a) MPI-MOM and (b) hybrid MPI/OpenMP-MOM. Maximum memory needed per core for (c) MPI-MOM and (d) hybrid MPI/OpenMP-MOM. Average wall clock time per iteration for (e) MPI-MOM and (f) hybrid MPI/OpenMP-MOM. Dashed lines are ideal scalability tangents.

Figure 6: Pictorial description of the matrix transpose during the FFTs at the propagation step. Slab decomposition before and after the transpose for the 3-D (a) FFTs and (b) inverse FFTs. The straight and dashed red lines show the slab assigned to each process before and after the transpose, respectively. The symbols $C_1 - C_6$ and $R_1 - R_6$ identify column slabs and row slabs, respectively.

Figure 7: 3-D FFTs using collective vs. point-to-point communication. Average wall clock time per iteration required during matrix solve step of the $L_\mathrm{P} = 8\lambda$ plate, $L_\mathrm{S} = 2\lambda$ sphere, and $L_\mathrm{A} = 6.8\lambda$ airplane simulations for (a) $T = 1$ and (b) $T = 4$.

Figure 8: Computational requirements for the $L_\mathrm{P} = 2\lambda$, $L_\mathrm{P} = 8\lambda$, and $L_\mathrm{P} = 32\lambda$ plate simulations using MPI-AIM and hybrid MPI/OpenMP-AIM. Wall clock time (matrix fill) for (a) MPI-AIM and (b) hybrid MPI/OpenMP-AIM. Average wall clock time per iteration (matrix solve) for (c) MPI-AIM and (d) hybrid MPI/OpenMP-AIM. Maximum memory needed per core for (e) MPI-AIM and (f) hybrid MPI/OpenMP-AIM. Dashed lines are ideal scalability tangents.

Figure 9: Average wall clock time per iteration (matrix solve): (a) Plate, (c) sphere, and (e) model airplane simulations for $T = 1$ and (b) plate, (d) sphere, and (f) model airplane simulations for $T = 4$.

Table 1: Parameters for scattering analysis of plates

| $L_\text{P}/\lambda$ | $N$ | $N^\text{C}$ | Reference | $err_{\theta\theta}(\%)$ |
|---|---|---|---|---|
| 1 | 280 | $12\times12\times4$ | MOM/AIM | 0.10/0.08 |
| 2 | 1 160 | $20\times20\times4$ | MOM/AIM | 0.36/0.34 |
| 4 | 4 720 | $36\times36\times4$ | MOM/AIM | 0.42/0.39 |
| 8 | 19 040 | $72\times72\times4$ | MOM/AIM | 0.34/0.31 |
| 16 | 76 840 | $144\times144\times4$ | MOM/AIM | 0.46/0.27 |
| 32 | 306 560 | $288\times288\times4$ | AIM | 0.30 |
| 64 | 1 227 520 | $576\times576\times4$ | AIM | 0.29 |
| 128 | 4 912 640 | $1152\times1152\times4$ | AIM | 0.29 |
| 256 | 19 655 680 | $2304\times2304\times4$ | AIM | 0.30 |

Table 2: Parameters for scattering analysis of spheres

| $L_\text{S}/\lambda$ | $N$ | $N^\text{C}$ | Reference | $err_{\theta\theta}(\%)$ |
|---|---|---|---|---|
| 1 | 3 384 | $24\times24\times24$ | Mie/MOM/AIM | 0.93/0.19/0.21 |
| 2 | 10 947 | $48\times48\times48$ | Mie/MOM/AIM | 0.97/0.09/0.01 |
| 4 | 44 595 | $64\times64\times64$ | Mie/MOM/AIM | 0.97/0.17/0.45 |
| 8 | 179 130 | $128\times128\times128$ | Mie/AIM | 0.97/0.37 |
| 16 | 742 059 | $256\times256\times256$ | Mie/AIM | 0.80/0.28 |
| 32 | 2 903 916 | $512\times512\times512$ | Mie/AIM | 0.97/0.33 |
| 64 | 11 601 048 | $1024\times1024\times1024$ | Mie/AIM | 0.98/0.31 |

Table 3: Parameters for scattering analysis of model airplanes

| $L_\text{A}/\lambda$ | $N$ | $N^\text{C}$ | Reference | $err_{\theta\theta}(\%)$ |
|---|---|---|---|---|
| 6.8 | 23217 | $75\times75\times32$ | MOM/AIM | 0.47/0.41 |
| 13.6 | 92868 | $144\times144\times48$ | MOM/AIM | 0.41/0.38 |
| 27.2 | 341472 | $256\times256\times96$ | AIM | 0.44 |
| 54.4 | 1485888 | $480\times480\times144$ | AIM | 0.47 |
| 108.8 | 5943552 | $960\times960\times288$ | AIM | 0.46 |

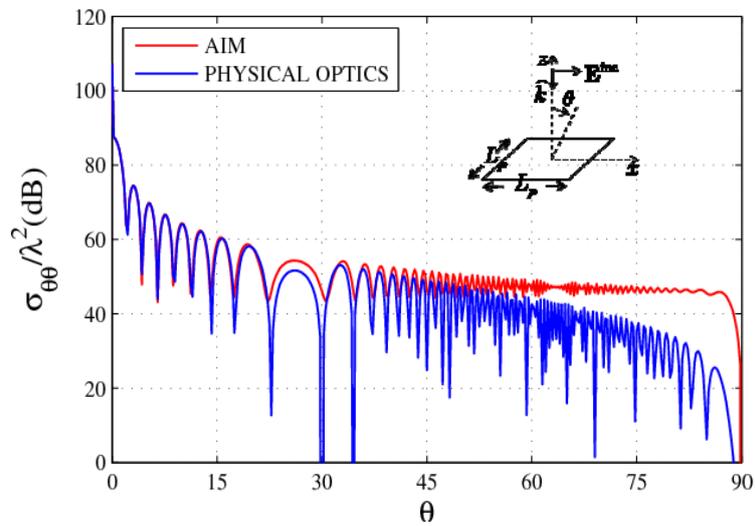
Figure 1(a)

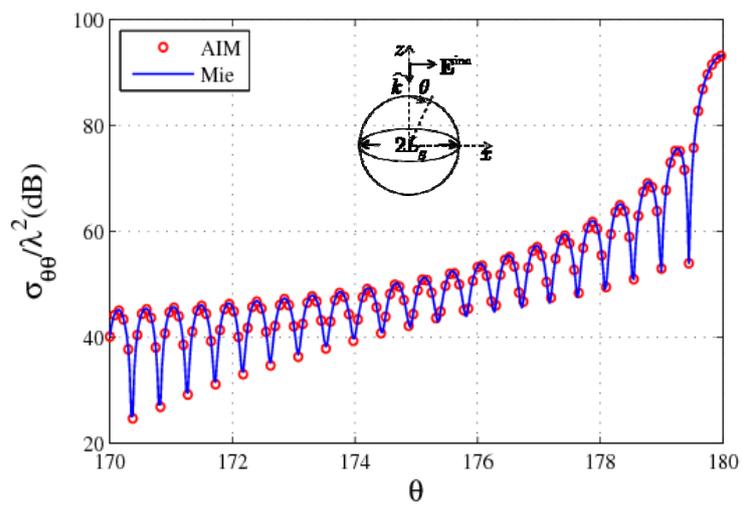
Figure 1(b)

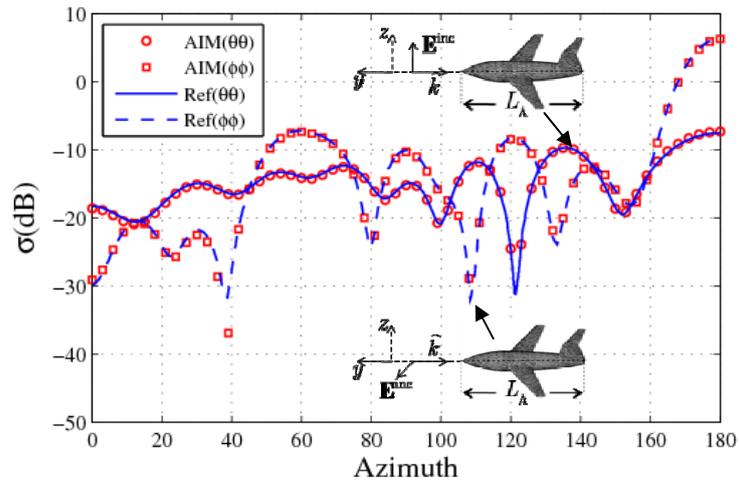

Figure 2

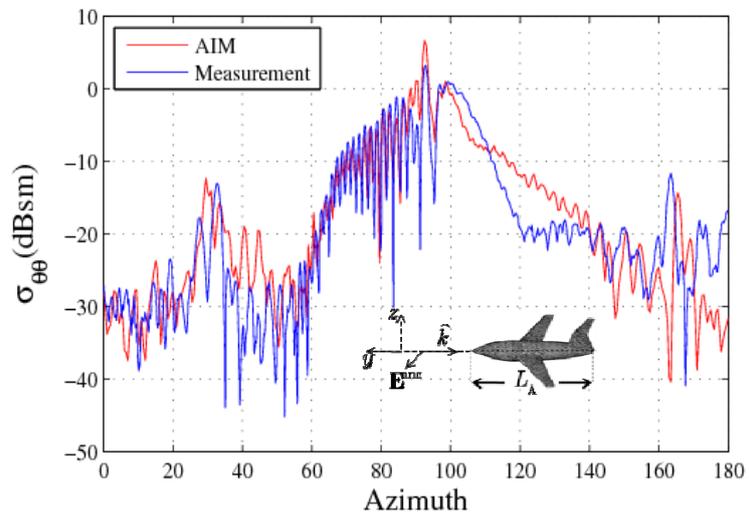

Figure 3

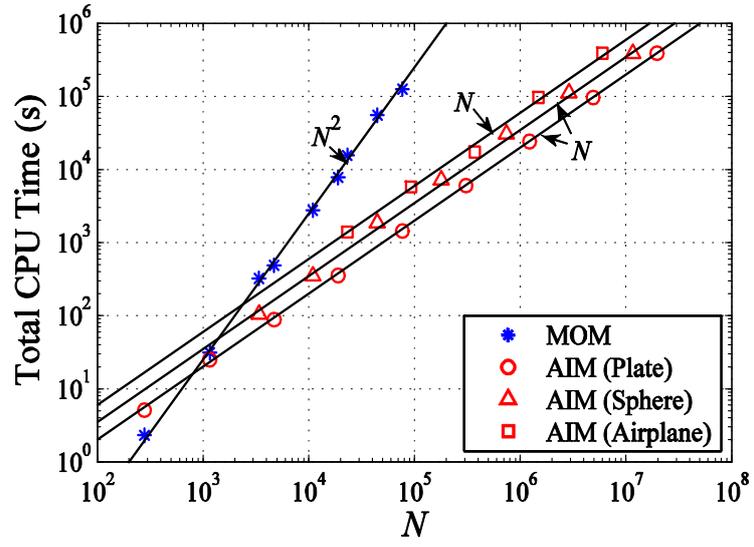

Figure 4(a)

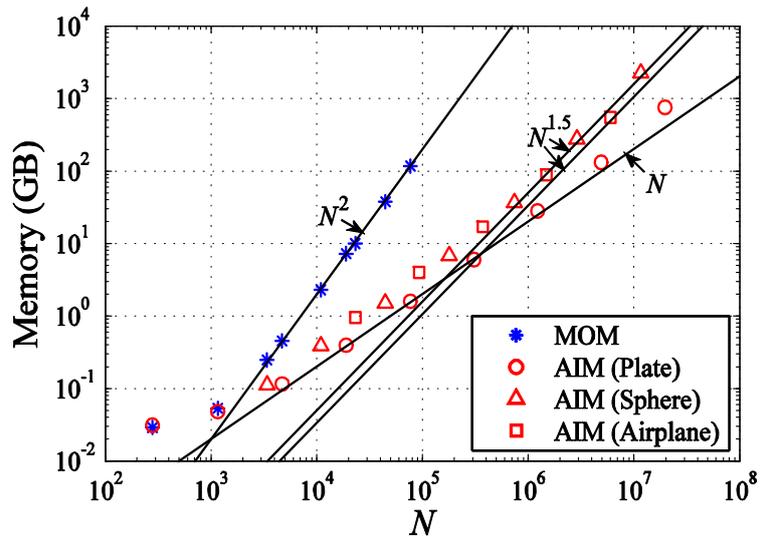

Figure 4(b)

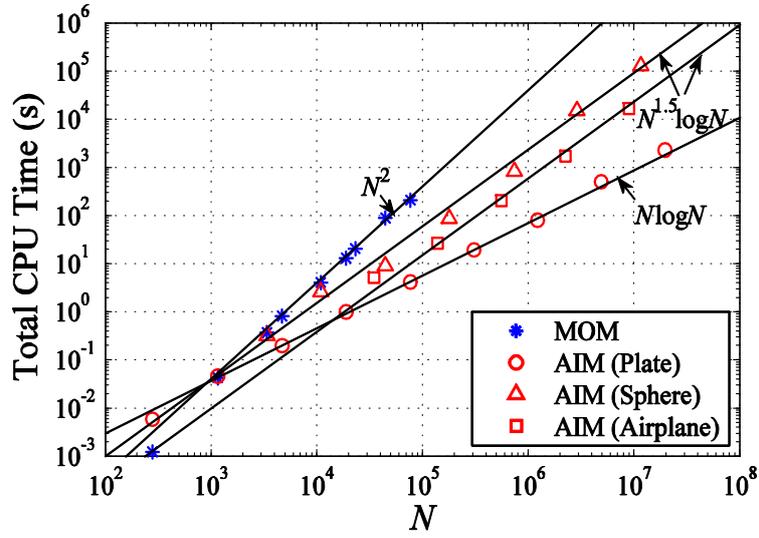

Figure 4(c)

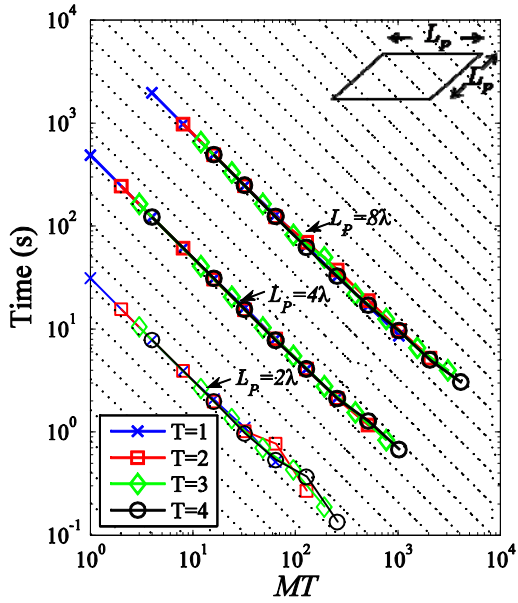

Figure 5(a)

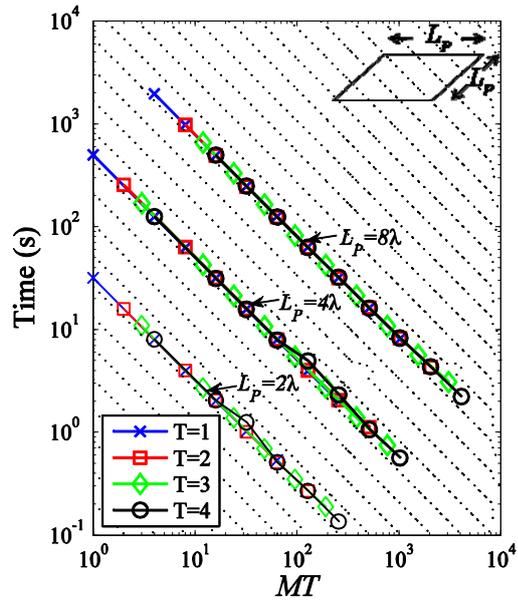

Figure 5(b)

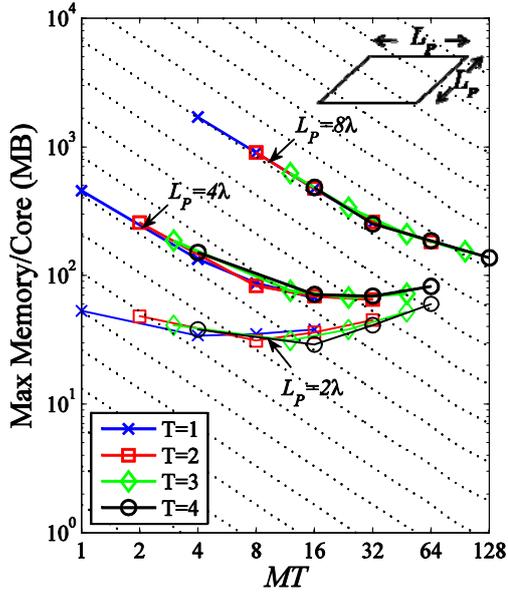

Figure 5(c)

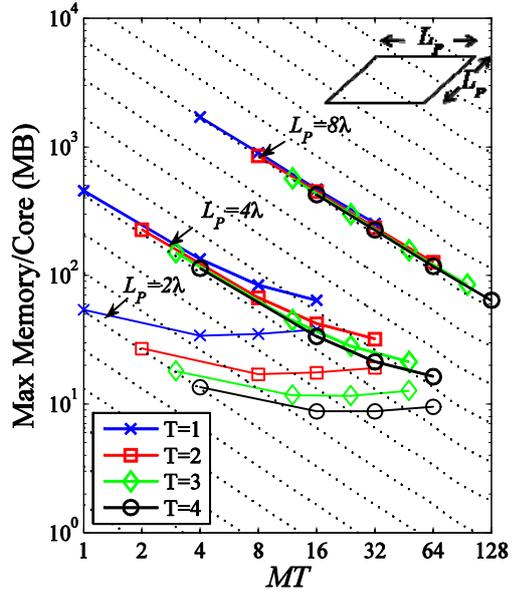

Figure 5(d)

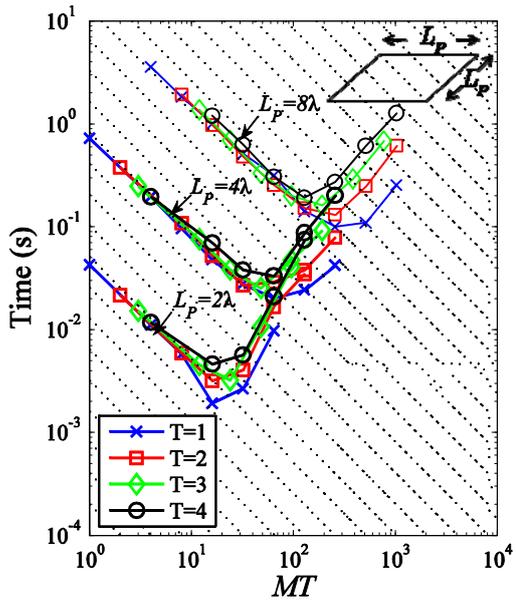

Figure 5(e)

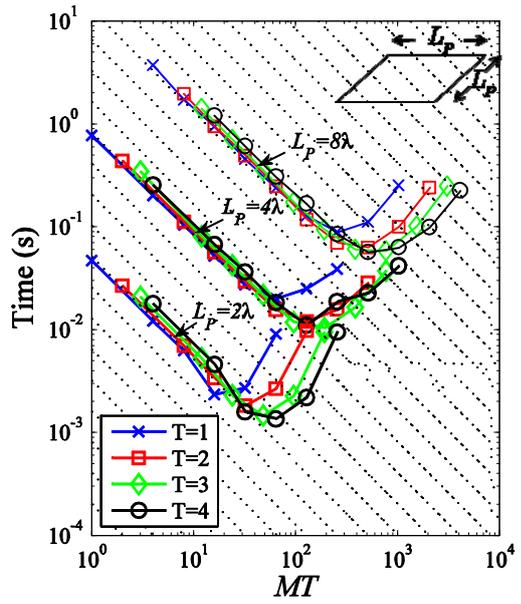

Figure 5(f)

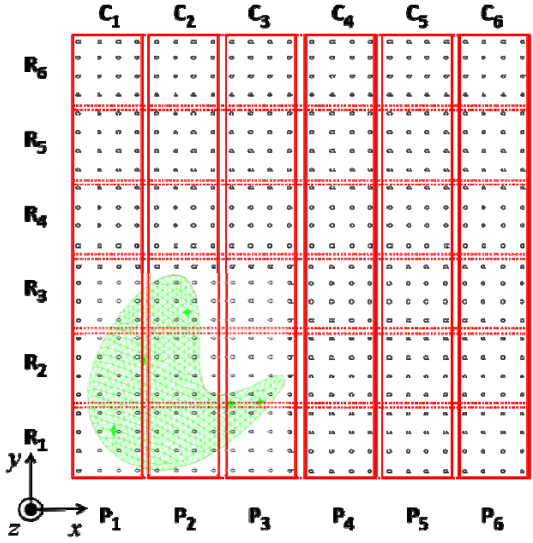
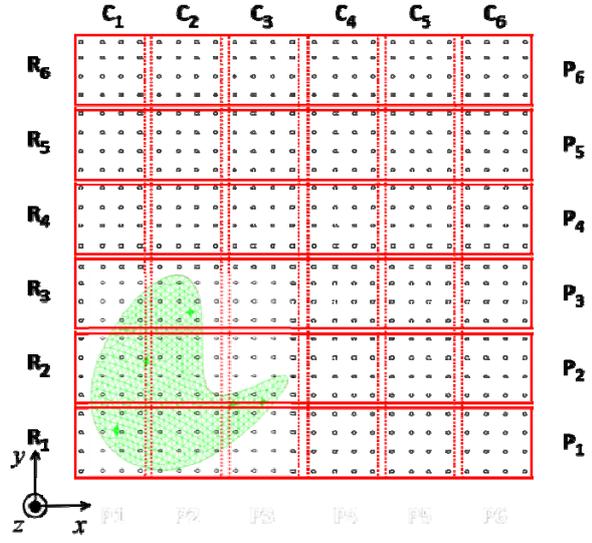

Figure 6(a)                                  Figure 6(b)

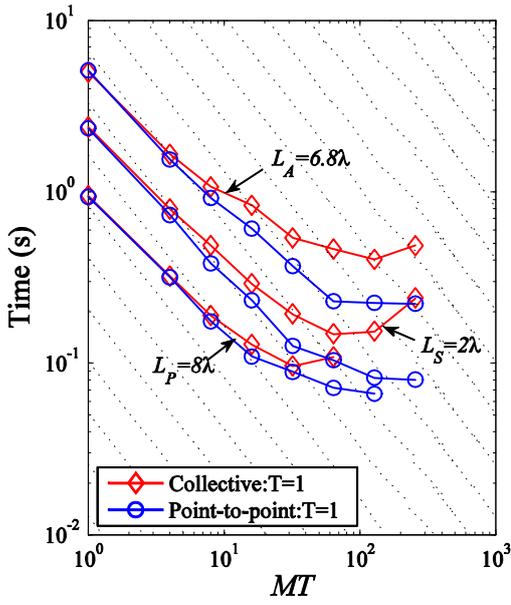
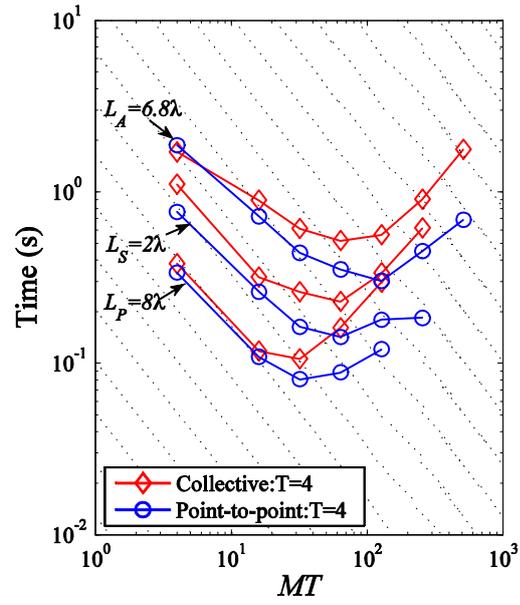

Figure 7(a)                                  Figure 7(b)

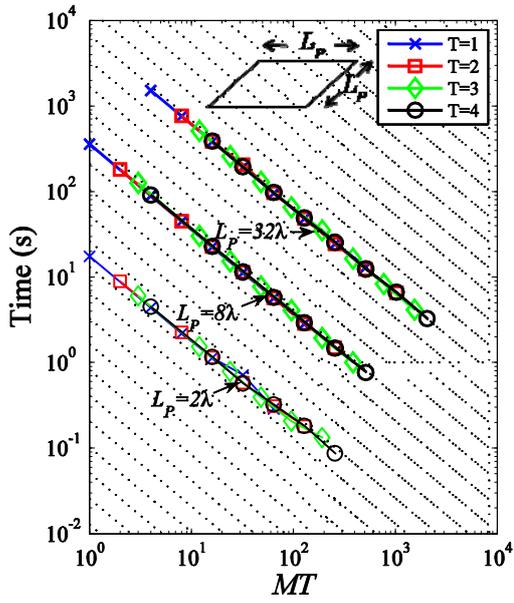

Figure 8(a)

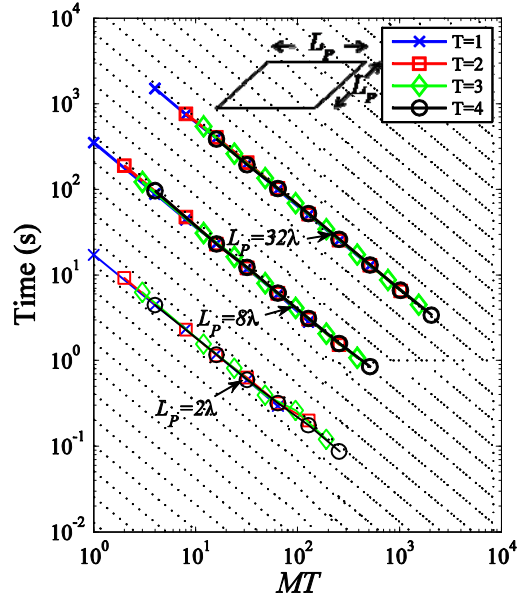

Figure 8(b)

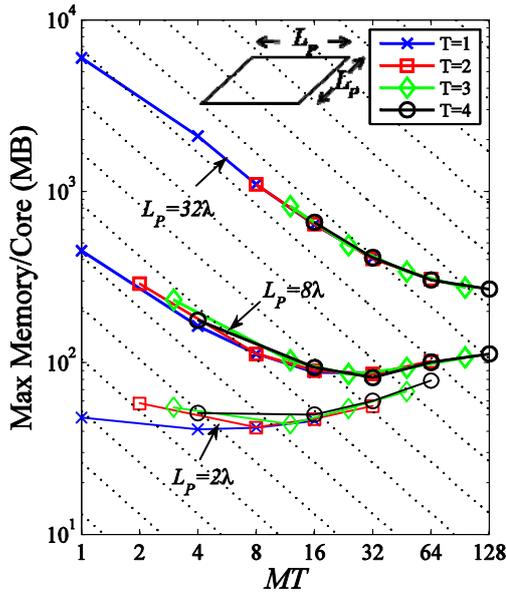

Figure 8(c)

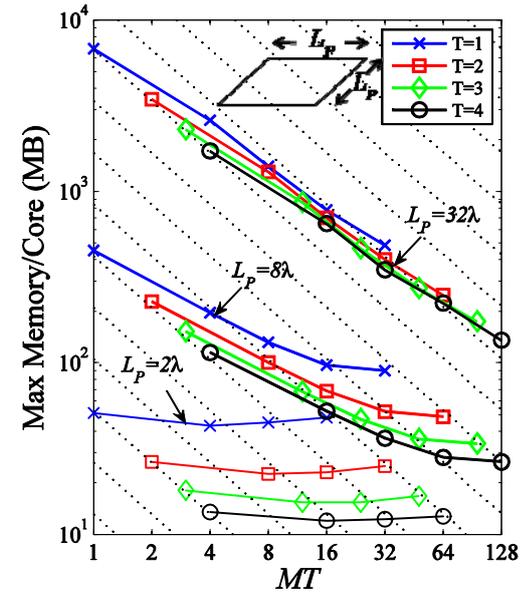

Figure 8(d)

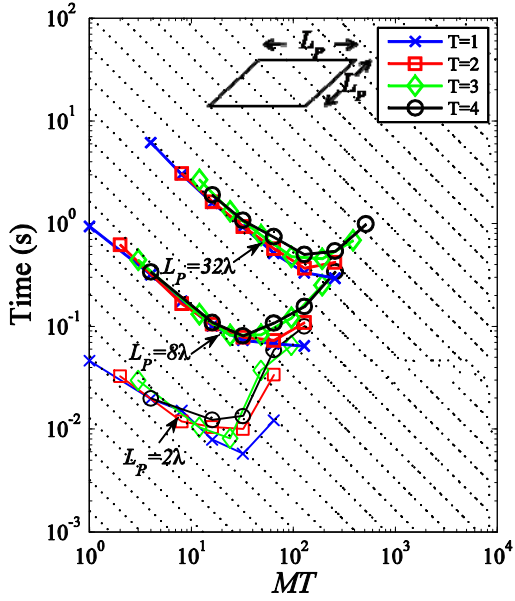

Figure 8(e)

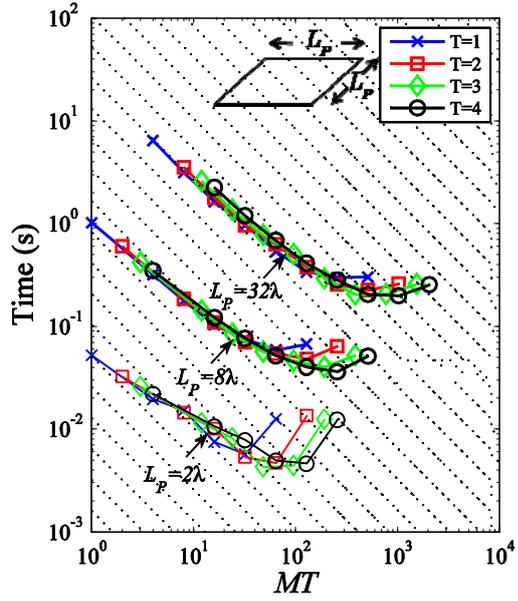

Figure 8(f)

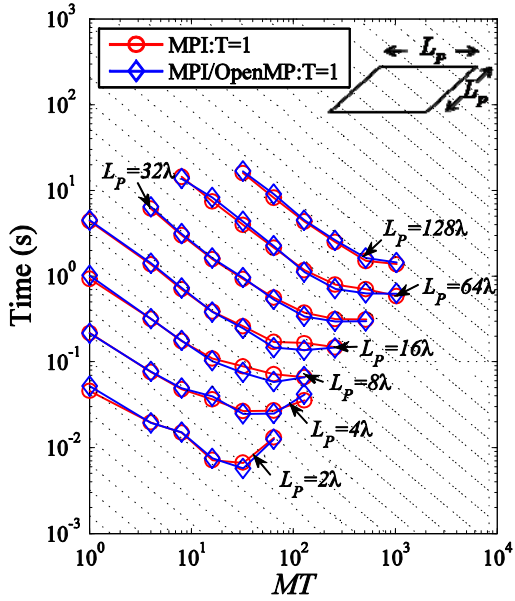

Figure 9(a)

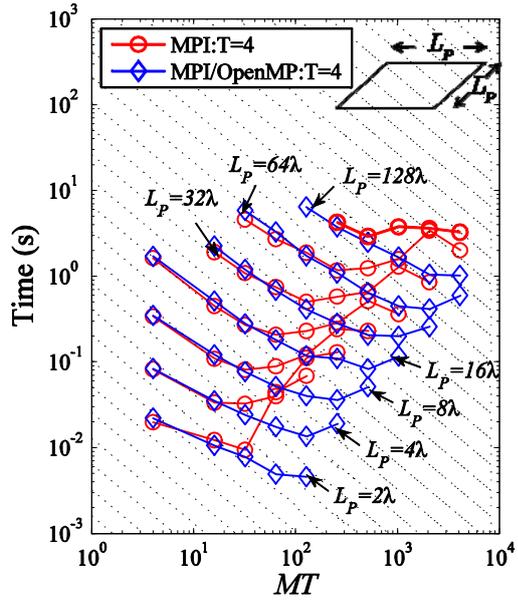

Figure 9(b)

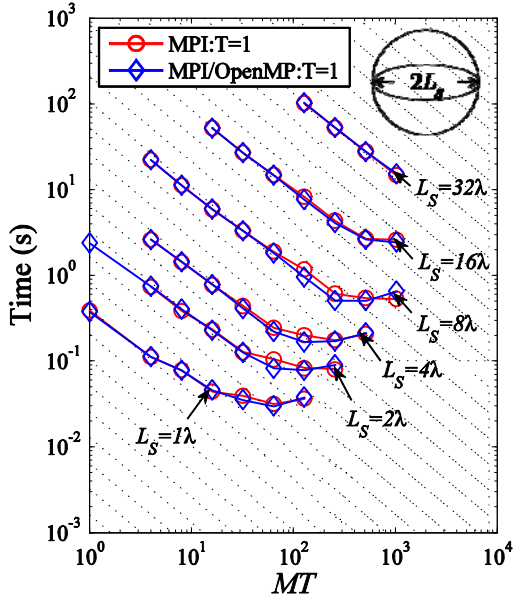

Figure 9(c)

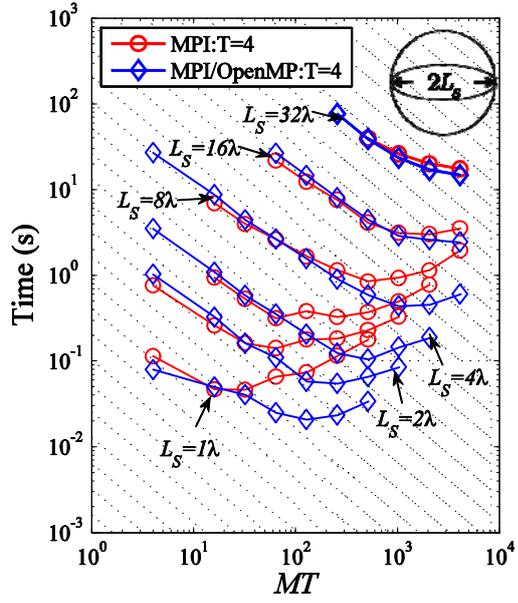

Figure 9(d)

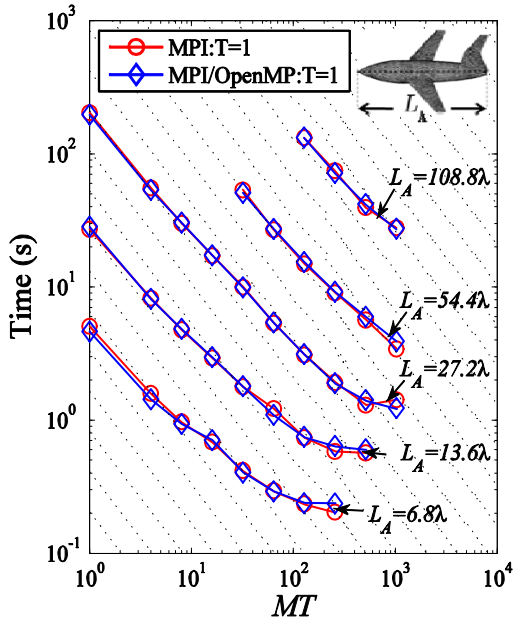

Figure 9(e)

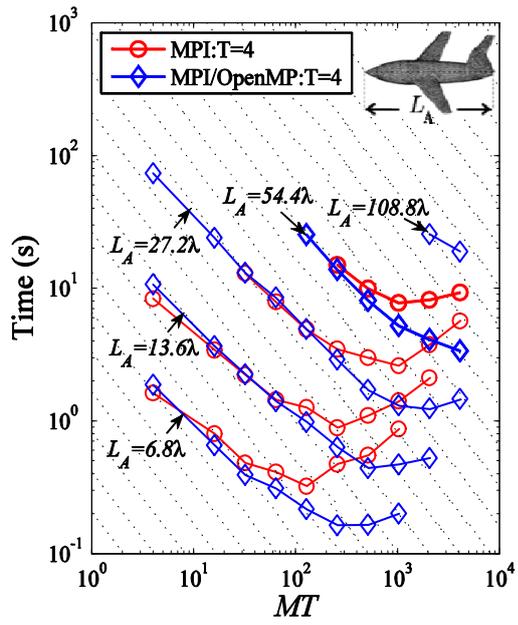

Figure 9(f)

24